\documentclass{article}
\usepackage{times,amsmath,amsfonts}

\newcommand{\smallspace}{\mskip 2mu minus 1mu}
\newcommand{\tinyspace}{\mskip 1mu}

\def\squareforqed{\hbox{\rlap{$\sqcap$}$\sqcup$}}
\def\qed{\ifmmode\squareforqed\else{\unskip\nobreak\hfil
\penalty50\hskip1em\null\nobreak\hfil\squareforqed
\parfillskip=0pt\finalhyphendemerits=0\endgraf}\fi}

\newtheorem{theorem}{Theorem}
\newtheorem{lemma}[theorem]{Lemma}
\newenvironment{proof}{\begin{trivlist}\item[]{\flushleft\bf Proof }}
{\qed\end{trivlist}}

\newcommand{\ket}[1]{\mbox{$| #1 \rangle$}}
\newcommand{\bra}[1]{\mbox{$\langle #1 |$}}
\newcommand{\inner}[2]{\mbox{$\langle #1 | #2 \rangle$}}

\DeclareMathSymbol{\leqslant}{\mathrel}{AMSa}{"36}
\newcommand{\subgroup}{\leqslant} 

\newcommand{\op}[1]{\mathsf{#1}}

\begin{document}

\title{The quantum query complexity of the
hidden subgroup problem is polynomial}

\date{January 12, 2004}

\author{
Mark Ettinger\\
{\protect\small\sl Los Alamos National Laboratory\/}%
\thanks{\,Email: \{\texttt{ettinger,knill}\}\texttt{$\mathchar"40$lanl.gov}.}
\and 
Peter H{\o}yer\\
{\protect\small\sl University of Calgary\/}%
\thanks{\,Email: \texttt{hoyer}\texttt{$\mathchar"40$cpsc.ucalgary.ca}.}
\and 
Emanuel Knill\\
{\protect\small\sl Los Alamos National Laboratory\/}\footnotemark[1]
}

\maketitle

%======================================================================

\begin{abstract}
We present a quantum algorithm which identifies with certainty a
hidden subgroup of an arbitrary finite group $G$ in only a polynomial
(in $\log |G|$) number of calls to the oracle.  This is exponentially
better than the best classical algorithm.  However our quantum
algorithm requires exponential time, as in the classical case.  Our
algorithm utilizes a new technique for constructing error-free
algorithms for non-decision problems on quantum computers.
\end{abstract}

%======================================================================

\section{Introduction}

Let $G$ be a finite group, written multiplicatively with
identity~$1_G$.  A~function~$f$ on~$G$ (with arbitrary range) is said
to be \emph{$H$-periodic} if $f$ is constant on the left cosets of a
subgroup~$H$ of~$G$.  If~$f$ also takes distinct values on distinct
cosets we say $f$ is \emph{strictly} $H$-periodic and we call $H$ the
\emph{hidden subgroup} of~$f$.  The \emph{hidden subgroup problem} (HSP) is
stated as follows: Given a description of~$G$ and a function~$f$ on~$G$
that is promised to be strictly $H$-periodic for some subgroup $H
\subgroup G$, find a generating set for~$H$.

Let $r$ denote the number of distinct subgroups of~$G$.  Fix any
ordering of the $r$ subgroups $(K_1, K_2, \ldots, K_r)$ satisfying
that $|K_{\mu}| \geq |K_{\mu+1}|$ for all $1 \leq
\mu < r$.  In~the HSP, we are searching for
a generating set for one out of~$r$ candidate subgroups.  Let $N=|G|$
denote the order of~$G$.
We consider $n=\log |G|$ to be the input size.
Since any subgroup of~$G$ is generated by a set of
at most $n$ elements of~$G$, the number~$r$ of distinct subgroups
of~$G$ is $2^{O(n^2)}$. 

We~assume the function~$f$ is given as an oracle so that the only way
we can gain knowledge about $f$ is by asking for its value on elements
of~$G$.  Formally, on~a quantum computer, the oracle is a unitary
operator $\op{O}_f$, that maps $\ket{g}\ket{0}$ to $\ket{g}\ket{f(g)}$
for all $g \in G$.
We assume without loss of generality that algorithms for the HSP
always output a subset of~$H$.  Suppose instead that an algorithm
outputs $X\not\subseteq H$. Then we can find the intersection of $X$
with $H$ by evaluating $f$ on each element $x\in X$ and only keeping
$x$ if $f(x)=f(1_G)$. This requires at most $|X|+1$ evaluations
of~$f$.

If~the group~$G$ is \emph{Abelian}, then it is possible to solve the
HSP in polynomial time with bounded error on a quantum computer.  That
is, we can efficiently find a subset $X \subseteq H$ that
generates~$H$ with probability at least~$\frac{2}{3}$.  This result
follows {from} the work of Simon~\cite{Simon}, Shor~\cite{Shor} and
Kitaev~\cite{Kitaev}.  It~is possible to improve the success
probability to one for Abelian groups of smooth order~\cite{BH} (a
group is of $c$-\emph{smooth order} if all prime factors of~$|G|$ are
at most $(\log |G|)^c$ for some constant~$c$).  For \emph{non-Abelian}
groups, our knowledge is much more
limited~\cite{EH,GSVV,HRT,IMS,RB,Zalka}.

The efficient HSP algorithm for Abelian groups of smooth order implies
that only a polynomial (in~$\log |G|$) number of calls to the oracle
are necessary to identify~$H$ with certainty.  The main result of this
paper is that this more limited result holds for all groups of finite
order.  That is, there exists a quantum algorithm that determines 
$H$ using a polynomial number of calls to the oracle. 

\begin{theorem}\label{thm:main}
There exists a quantum algorithm that, given a finite group~$G$ and an
oracle~$f$ on~$G$ promised to be strictly $H$-periodic for some
subgroup $H \subgroup G$, calls the oracle $O(\log^4|G|)$ times and
outputs a generating set for~$H$.  The algorithm fails with
probability exponentially small in $\log |G|$.  The algorithm can be
made exact in any model allowing arbitrary one-qubit gates.
\end{theorem}

An important consequence of this result is that it rules out most
known methods for proving super-polynomial lower bounds on
the total complexity of bounded-error quantum algorithms for the HSP.
Most of these methods bound the query complexity, including the recent
ones by Aaronson~\cite{Aaronson} and Shi~\cite{Shi}. This works well
for problems where the query complexity is at most
poly-logarithmically smaller than the time complexity. Because of our
result, one cannot obtain super-polynomial lower bounds on the total
complexity of algorithms for the HSP by bounding the query complexity.

Our result extends to exact quantum algorithms (algorithms that
determine the answer with certainty) in any model that allows
arbitrary one-qubit gates.  If allowing only a restricted set of
one-qubit gates, our work leaves a hope that one may be able to prove
a super-polynomial lower bound on the query complexity for the
\emph{exact} case.  

A proof of the upper bound on the query complexity only requires
establishing the existence of a sufficiently short sequence of unitary
operations and oracle calls on appropriately chosen quantum
systems. The sequence depends on the group.  Our proof explicitly
constructs the sequence and makes it apparent how to realize the
unitary operations using quantum gates {from} a universal set.  In fact,
the sequence can be obtained by means of a (classical) preprocessing
algorithm with input a specification of $G$ and whose output is the
required sequence of gates and oracle calls. For solving the HSP
exactly, the classical preprocessing algorithm requires exact real
number arithmetic and access to trigonometric functions of rational
angles. The preprocessing algorithms and the quantum networks they
compute are inefficient.

%======================================================================
\section{The algorithm}

Our~proof of Theorem~\ref{thm:main} consists of two stages.  
In subsection~\ref{subsec:error},
we give a
quantum algorithm that identifies the correct subgroup with
exponentially small error probability, and 
in subsection~\ref{subsec:exact},
we then show how to reduce
the error probability to zero. 
We begin with an overview.

We~use $2+2s$ registers, where $s$ is a positive integer
that will be chosen to achieve sufficiently low error probability.
The first register is the output register and contains
an integer~$\nu$ (a subgroup index) between $0$ and~$r$.
The second register is used as a counter and contains
an integer~$\ell$ between $0$ and~$r$.
The remaining $2s$ registers are grouped in $s$ blocks,
each consisting of $2$~consecutive registers (a ``couplet'') called 
the ``subgroup'' and the ``function'' register.
Within each couplet, the first register contains an element of~$G$ and
the second a value in the range of~$f$.

We start by creating the initial state
\begin{equation}
\ket{\Psi_{\textup{init}}}
= \ket{0}\ket{0} \otimes
\Bigg(\frac{1}{\sqrt N} \sum_{g \in G}
\ket{g}\ket{f(g)}\Bigg)^{\otimes s}.
\end{equation}
This superposition can be created efficiently using $s$ applications
of operator~$\op{O}_f$.  
We then apply the unitary operator ${\op{Test}}$, to be defined in
subsection~\ref{subsec:error}, producing the superposition
$\ket{\Psi_{\textup{final}}} = {\op{Test}}
\ket{\Psi_{\textup{init}}}$.  We measure the first register of
$\ket{\Psi_{\textup{final}}}$, yielding some subgroup-index $\nu$ as
outcome.  If $1\leq \nu\leq r$, we output a generating set
for~$K_{\nu}$, otherwise we output~$\{1_G\}$, which may
be the wrong answer.
Our algorithm has
exponentially small error probability.

\begin{theorem}\label{thm:error}
Let $\textup{Prob}[K_{\nu}|H]$ denote the probability that the outcome
of the measurement of the first register of $\ket{\Psi_{\textup{final}}}$
is~$\nu$, conditioned on the hidden subgroup being~$H$.
Then $\textup{Prob}[H|H] \geq 1 - 4r/2^{s/2}$ for all
subgroups~$H \subgroup G$, where $r$ is the number of subgroups of~$G$
and $s$ is the number of queries.
In particular, for $s \in \Theta(n^2 + \log (1/\epsilon))$, 
the algorithm outputs the correct subgroup with probability at least
$1-1/\epsilon$.
\end{theorem}

The theorem is proved in subsection~\ref{subsec:error}, and
in~subsection~\ref{subsec:exact}, we make this algorithm exact by
precomputing $\textup{Prob}[K|H]$ for each subgroup pair $(K, H)$,
adjusting the conditional probabilities to make them more uniform and
applying amplitude amplification~\cite{BHMT}.

%======================================================================
\subsection{An algorithm with exponentially small error probability}
\label{subsec:error}

A~(left) \emph{translation} for a
subgroup $K$ of~$G$ is a subset~$T \subseteq G$ so that any element $g
\in G$ can be written uniquely in the form $g = tk$ for some $t \in T$
and $k \in K$. Fix a translation $T_{\mu}$ for each of
the $r$ subgroups $K_{\mu}$ of~$G$.

The operator ${\op{Test}}$ tests the hidden subgroup
for each of the $r$ candidate subgroups, one by one.  It is
defined by
\begin{equation}
{\op{Test}} \;=\;
{\op{Test}}_{r} \cdot\smallspace \cdots\smallspace \cdot
{\op{Test}}_{2} \cdot
{\op{Test}}_{1},
\end{equation}
where each ${\op{Test}}_{\mu}$ is a unitary operator that tests
whether $f$ is $K_\mu$-periodic.  If~a function is $K$-periodic, it is
also $K'$-periodic for any proper subgroup $K'$ of~$K$, so we test
for bigger subgroups first by requiring that $|K_{\mu}| \geq
|K_{\mu+1}|$ for all $1 \leq \mu < r$.
When we find that $f$ is $K_{\mu}$-periodic for some subgroup
$K_\mu$, we record this in the first register, and we begin counting
in the second register.
For every subgroup $K_{\mu} \subgroup G$, let $\op{Q}_{\mu}$ be any
unitary operator acting on the first two registers that 
satisfies 
\begin{equation*}
\op{Q}_{\mu}: \quad \left\{
\begin{array}{rcl}
\ket{0}\ket{0} & \mapsto & \ket{\mu}\ket{1} \\
\ket{\nu}\ket{\ell} & \mapsto & \ket{\nu}\ket{\ell+1}, 
\qquad\text{ if $\ell>0$.}
\end{array}
\right.
\end{equation*}
Once the count $\ell$ in the second register is increased {from} its
initial value of~$0$ to~$1$, the contents of the first register are
never changed.  The purpose of the counter is to ensure
unitarity and that once some test succeeds, no future test
affects the contents of the first register.

We test for $K_{\mu}$-periodicity by acting on the $s$ couplets.
If function $f$ is $K_{\mu}$-periodic then the $s$ subgroup registers
are in a superposition of the coset states
$\ket{tK_{\mu}}
  = {\frac{1}{\sqrt{|\smash[b]{K_{\mu}}|}}}
    \sum_{k\in K_{\mu}}\ket{tk}$.
Let ${\op{P}}_{s,\mu}$ be the projector of the $s$ couplets
defined by
\begin{equation*}
{\op{P}}_{s,\mu} = \Bigg(\sum_{t \in T_{\mu}} 
  \ket{t K_{\mu}}\bra{t K_{\mu}} \otimes \op{I}
\Bigg)^{\otimes s},
\end{equation*}
where $\op{I}$ denotes the identity operator,
and let ${\op{P}}_{s,\mu}^{\smash{\perp}}$ denote its complement.
Define operator~${\op{Test}}_{\mu}$ by
\begin{equation}\label{eq:testmu}
{\op{Test}}_{\mu} =
   \op{Q}_{\mu} \otimes {\op{P}}_{s,\mu}
 + \op{I} \otimes {\op{P}}^{\perp}_{s,\mu},
\end{equation}
which is unitary by construction.  Its effect is an application of
$\op{Q}_{\mu}$ on the first two registers, conditioned on having the
$s$ subgroup registers in coset states of $K_{\mu}$.  The condition
can be implemented with the help of any pair of unitary operators
$\op{U}_{\mu}$ and $\op{V}_{\mu}$, 
where $\op{U}_{\mu}$ maps $\ket{1_G}$ to
$\ket{K_{\mu}}$ and $\op{V}_{\mu}$ maps $\ket{t}\ket{k}$ to
$\ket{1_G}\ket{tk}$ for all $t \in T_{\mu}$ and $k\in K_{\mu}$.  The
procedure is as follows: Adjoin an ancilla register to each subgroup
register and apply $\op{V}_{\mu}^\dagger$ to these $s$ register 
pairs. Then
apply $\op{U}_{\mu}^\dagger$ to the subgroup registers. Next, coherently
apply $\op{Q}_{\mu}$ if all subgroup registers are in $\ket{1_G}$ and
finally reverse the previous steps.  It is possible to realize each of
these steps with a network of gates of complexity polynomial in $N$
and~$s$.

\begin{lemma}\label{lm:goodtest}
If~$f$ is $K_{\mu}$-periodic, then
\begin{equation*}
{\op{Test}}_{\mu} \ket{\Psi_{\textup{init}}}
\,=\, \ket{\mu}\ket{1} \otimes
\Bigg(\frac{1}{\sqrt{N}} \sum_{g \in G}
\ket{g}\ket{f(g)}\Bigg)^{\otimes s}.
\end{equation*}
\end{lemma}

\begin{proof}
We~assume in the lemma that $f$ is $K_{\mu}$-periodic,
that is, $f(t)=f(tk)$
for all $t \in T_{\mu}$ and $k \in K_{\mu}$, and hence the state 
$\frac{1}{\sqrt{N}} \sum_{g \in G} \ket{g}\ket{f(g)}
= \frac{1}{\sqrt{N}} \sum_{t \in K_{\mu}} \ket{tK_{\mu}}\ket{f(t)}$
is in the $+1$-eigenspace of $\op{P}_{1,\mu}$.
It follows that $\op{P}_{s,\mu}$ acts as the identity on the $s$ couplets, 
and thus applying operator ${\op{Test}}_{\mu}$ as defined
in Eq.~\ref{eq:testmu} on the initial state
$\ket{\Psi_{\textup{init}}}$
yields the state given on the right hand side in the equation of the
lemma.
\end{proof}

Since we iterate through $r$~tests, we require that if $f$ is
not $K_{\mu}$-periodic, then the state is so marginally altered that
it is safe to continue to test for $K_{\mu+1}$-periodicity.

\begin{lemma}
If~$f$ is not $K_{\mu}$-periodic, then the distance
$\big|({\op{Test}}_{\mu} \ket{\Psi_{\textup{init}}})
   - \ket{\Psi_{\textup{init}}}\big|$
is at most $\frac{2}{2^{s/2}}$.
\end{lemma}

\begin{proof}
Let $H$ denote the hidden subgroup of~$f$.
Consider the case $s=1$. Then
\begin{multline*}
|{\op{P}}_{s,\mu}\ket{H}\ket{f(H)}|^2 =
   \sum_{t\in T_{\mu}} |\inner{tK_{\mu}}{H}|^2 
  = \sum_{t\in T_{\mu}: tK_{\mu} \cap H\not=\emptyset}
       |\inner{tK_{\mu}}{H}|^2\\
  = (|H|/|K_{\mu}\cap H|)
       |K_{\mu}\cap H|^2/(|K_{\mu}||H|)
  = |K_{\mu} \cap H| / |K_{\mu}| \leq \frac{1}{2}.
\end{multline*}
It follows that for arbitrary $s$, the amplitude squared of 
$(\op{Q}_{\mu} \otimes {\op{P}}_{s,\mu})\ket{\Psi_{\textup{init}}}$
is upper bounded by $(\frac{1}{2})^s$.
Since ${\op{Test}}_{\mu}$ acts trivially on the orthogonal component
$(\op{I} \otimes {\op{P}}^\perp_{s,\mu})
  \ket{\Psi_{\textup{init}}}$,
the result follows.
\end{proof}

For each $1 \leq j \leq r$, let
$\ket{\Psi_{j}} =
{\op{Test}}_{j} \cdot\smallspace \cdots\smallspace \cdot
{\op{Test}}_{1} \ket{\Psi_{\textup{init}}}$
denote the state of the system after $j$ tests.
By~the above lemma, it is safe to iterate through all tests,
since distances can add up only linearly.

\begin{lemma}\label{lm:linearly}
If~$f$ is not $K_{\mu}$-periodic
for any $1 \leq \mu \leq j$, then the distance
$\big|\ket{\Psi_{j}} - \ket{\Psi_{\textup{init}}}\big|$
is at most~$\frac{2j}{2^{s/2}}$.
\end{lemma}

Suppose that the input function $f$ is strictly $K_{\nu}$-periodic.
Then, by Lemma~\ref{lm:linearly},
the state $\ket{\Psi_{\nu-1}}$
just prior the test ${\op{Test}}_{\nu}$ is at most at a distance
$\epsilon = \frac{2r}{2^{s/2}}$
away {from} the initial state $\ket{\Psi_{\text{init}}}$.
Thus the probability
that test ${\op{Test}}_{\nu}$ fails in producing the
correct answer $\ket{\nu}$ in the first register is
at most~$2\epsilon = \frac{4r}{2^{s/2}}$ by Lemma~\ref{lm:goodtest}.

We note that operator ${\op{Test}}$ never acts on the $s$ function
registers.  One can therefore measure these prior to the application
of ${\op{Test}}$ without affecting the error probability of the
bounded error algorithm.  However, our exact algorithm requires
unitarity and assumes that the function registers are not measured.

The probability of measuring the outcome~$\mu$ depends on which
subgroup $H$ is the hidden subgroup, but it is independent of the
values $f$ takes on the different cosets of~$H$.  That is, for any two
functions $f$ and~$f'$ having the same hidden subgroup~$H$, the
probabilities of measuring~$\mu$ are the same.  We may therefore let
$\textup{Prob}[K_{\mu}|H]$ denote the probability that $\mu$
is the outcome of measuring the first register 
of $\ket{\Psi_{\textup{final}}}$, conditioned on the hidden 
subgroup being~$H$.
Theorem~\ref{thm:error} follows.

%======================================================================
\subsection{An exact algorithm}
\label{subsec:exact}

We next use amplitude amplification to make our algorithm exact.  
This requires the ability to compute exactly the conditional
probabilities $\textup{Prob}[K_{\mu}|H]$ without using the oracle.  One
method for computing $\textup{Prob}[K_{\mu}|H]$ is to pick an
arbitrary function $f$ that is strictly $H$-periodic, and simulate the
quantum computation of ${\op{Test}}$ on oracle~$f$ with a classical
computer. Note that the classical computer implements arithmetic on
exact real numbers. 
However, neither this nor the high
complexity of the algorithm is relevant to our proof of low query
complexity. For this purpose we only need to know that the appropriate
unitary transformations between queries exist.  Thus, our quantum
algorithm runs in exponential time, but uses only polynomially many
queries in any model allowing arbitrary one-qubit and two-qubit gates,
where each gate is given (implicitly) via the result of a classical
computation.  This model is of course not realistic, but it suffices
to rule out easy query complexity lower bounds, as discussed in the
Introduction.

Let $Y_{\scriptscriptstyle \frac{1}{4}} \cup Y_{\scriptscriptstyle
\frac{3}{4}}$ be any partitioning of the set of subgroups
$\{K_1,\ldots,K_r\}$.  The algorithm ${\op{Test}}$ of the previous
subsection succeeds in identifying the hidden subgroup with high
probability.  We now describe a new algorithm ${\op{ExactTest}}$ that
merely distinguishes between the two above sets of subgroups, but does
so with known and desirable probabilities.

\begin{lemma}\label{lm:34}
The probability that the outcome of a measurement of the ancilla qubit
of the state
${\op{ExactTest}}(\ket{\Psi_{\textup{init}}} \otimes \ket{0})$
is~1 is $\frac34$ if the hidden subgroup $H$ is in
$Y_{\scriptscriptstyle \frac{3}{4}}$,
and it is $\frac14$ if $H$ is in $Y_{\scriptscriptstyle \frac{1}{4}}$.
\end{lemma}

Before we describe the algorithm ${\op{ExactTest}}$,
let $M$ be an $r \times r$ matrix over $[0,1]$
with each row and column indexed by a subgroup.
Let entry $(H,K_{\mu})$ of~$M$ be the conditional probability
$\textup{Prob}[K_{\mu}|H]$ that a
measurement of the first register of~$\ket{\Psi_{\textup{final}}}$
yields the outcome $\mu$
conditional on  $f$ being strictly $H$-periodic.

Let $s = \lceil2 \log(4 r^3)\rceil \in O(\log^2 N)$ so that
by~Theorem~\ref{thm:error}, any~diagonal entry of~$M$ is at least
$1-\frac{1}{r^2}$, and since the entries of any row of $M$ sum to~1,
any off-diagonal entry of $M$ is between $0$ and~$\frac{1}{r^2}$.
Thus, we can express $M$ as $M = I - \Delta$, where each entry of
$\Delta$ has absolute value bounded by $\frac{1}{r^2}$.  
It follows that $M^{-1} = I + \Delta + \Delta^2 + \Delta^3 + \cdots$,
subject to the convergence of $\Gamma = \Delta + \Delta^2 + \Delta^3 +
\cdots$, which we now show.  By induction on $i$, each entry of
$\Delta^i$ has absolute value bounded by $\frac{1}{r^{i+1}}$.
Therefore, each entry of $\Gamma$ has absolute value bounded by
$\sum_{i=1}^{\infty} \frac{1}{r^{i+1}} = \frac{1}{r(r-1)}$.

Let $y$ be any $r \times 1$ column vector with entries {from}
$\{\frac{1}{4}, \frac{3}{4}\}$ and with each row indexed by a
subgroup.  Set $x = M^{-1} y$.  Then, since $M^{-1} y = y + \Gamma y$,
every entry of~$x$ is within $\frac{3}{4(r-1)}$ of the corresponding
entry of $y$, and thus every entry of~$x$ is in $[0,1]$ for $r \geq
4$.

Algorithm ${\op{ExactTest}}$ acts on the initial state
$\ket{\Psi_{\textup{init}}} \otimes \ket{0}$, where the last register
holds an ancilla qubit in state~$\ket{0}$, and is defined as
\begin{equation}
{\op{ExactTest}} = \op{R} \cdot (\op{Test} \otimes \op{I}).
\end{equation}
First, it applies $\op{Test}$ on the first part of the system.  
It then applies $\op{R}$, which, conditionally on the output register
holding the subgroup index $\mu$, rotates the ancilla qubit {from}
$\ket{0}$ to $\sqrt{1-\smash{x_\mu}}\ket{0} +
\sqrt{\smash{x_\mu}\vphantom{1}}\ket{1}$.  Because $\op{R}=\sum_\mu
\op{P}_\mu \otimes \op{R}_\mu$ for projectors
$\op{P}_\mu=\ket{\mu}\bra{\mu}$ and certain qubit rotations
$\op{R}_\mu$, it can be implemented unitarily.  The probability that a
measurement of the ancilla qubit of the resulting state
${\op{ExactTest}}\big(\ket{\Psi_{\textup{init}}} \otimes \ket{0}\big)$
yields a~1 is thus 
\begin{equation*}
\sum_{\mu}
  x_{\mu} \tinyspace\textup{Prob}[K_{\mu}|H_{\nu}],
\end{equation*}
which, by definition of the column vector~$x$,
is equal to $y_{\nu}$, where $H_{\nu}$ is the hidden subgroup.
In other words, the probability of measuring a~1 depends only on
the index of the hidden subgroup.
Set $y_{\nu}=\frac34$ if $K_{\nu} \in
Y_{\scriptscriptstyle \frac{3}{4}}$, and
set $y_{\nu}=\frac14$ if $K_{\nu} \in
Y_{\scriptscriptstyle \frac{1}{4}}$.
Lemma~\ref{lm:34} follows.

Lemma~\ref{lm:34} provides us with a method for distinguishing between
two complementary subsets of subgroups with probabilities $\frac34$
and~$\frac14$.  Using amplitude amplification~\cite{BHMT}, we can
alter those probabilities into being equal to $0$ and~$1$, and hence
distinguish between the two sets $Y_{\scriptscriptstyle \frac{3}{4}}$
and $Y_{\scriptscriptstyle \frac{1}{4}}$ with certainty.  Applying
binary search on the set of subgroups with various choices of
$Y_{\scriptscriptstyle \frac{3}{4}}$ and $Y_{\scriptscriptstyle
\frac{1}{4}}$ then yields the second half of Theorem~\ref{thm:main}.

%======================================================================
\section{Concluding remarks}

Let $\textsf{HSP}$ denote the decision problem of determining if the
hidden subgroup is non-trivial.  Let $Q_E(\mathcal{P})$ denote the
quantum \emph{query} complexity of determining some decision
problem~$\mathcal{P}$ with certainty, and let $Q_1(\mathcal{P})$
denote the quantum query complexity of determining~$\mathcal{P}$ with
one-sided error.  Then $Q_E(\textsf{HSP}) \in O(\log^2 |G|)$ since it
suffices to use one round of amplitude amplification, because binary
search among subgroups is not needed (let $Y_{\scriptscriptstyle
\frac{3}{4}}$ be the singleton containing only the trivial subgroup).
Also $Q_1(\textsf{HSP}) \in O(\log |G|)$ since we need to test only
for the cyclic subgroups, of which there are at most~$|G|$, and then
use one round of amplitude amplification for the case where the
subgroup is trivial.  If the subgroup is non-trivial, the algorithm
may output an incorrect answer and thus the algorithm has one-sided
error.

The technique to construct exact quantum algorithms presented here
relies on the property that we can compute the conditional
probabilities with arbitrary precision.  The technique seems to be
applicable both to proving lower bounds as well as to designing
efficient algorithms. It can rule out easy lower bounds for exact
quantum computation, or it can be used to give simple and efficient
exact quantum algorithms for problems for which the number of distinct
success probabilities is polynomially bounded in the running time of 
the given bounded-error algorithm.

\section*{Acknowledgements}
We are grateful to Richard Cleve for suggesting studying the
exact query complexity of the HSP and for 
valuable comments and encouragement.  We appreciate the
constructive comments of the referees.  M.~E.~and E.~K.\ were supported
by the DOE, contract W-7405-ENG-36, and by the
NSA.  P.~H.~received support {from} Alberta Ingenuity Fund and the
Pacific Institute for the Mathematical Sciences.

%======================================================================
% \bibliographystyle{elsart-num}

\end{document}